\documentclass{article}

\usepackage[dblblindworkshop,final,nonatbib]{neurips_2026}

\usepackage[utf8]{inputenc} 
\usepackage[T1]{fontenc}    
\usepackage{hyperref}       
\usepackage{url}            
\usepackage{amsfonts}       
\usepackage{microtype}      
\usepackage{xcolor}         
\usepackage{amsmath}        
\usepackage{graphicx}       
\usepackage{subcaption}     
\usepackage{cleveref}
\usepackage[style=numeric-comp,sorting=none]{biblatex}
\usepackage{enumitem}

\title{Reconfigurable On-Chip AI for Particle Physics Detectors with Logic Neural Networks}
\workshoptitle{On-Device Intelligence: Foundation Models under Real-World Constraints}

\author{%
  Mila Bileska \\
  Princeton University \\
  \texttt{mbileska@princeton.edu} \\
  \And
  Lino Gerlach \\
  Princeton University\\
  \texttt{lg0508@princeton.edu} \\
  \And
  Jennet Dickinson \\
  Cornell University \\
  \texttt{jennet.dickinson@cornell.edu} \\
  \And
  Lindsey Gray \\
  Fermilab \\
  \texttt{lagray@fnal.gov} \\
  \And
  Giuseppe Di Guglielmo \\
  Fermilab \\
  \texttt{gdg@fnal.gov}
  \And
  Isobel Ojalvo \\
  Princeton University\\
  \texttt{iojalvo@princeton.edu} \\
  \And
  Peter Elmer \\
  Princeton University \\
  \texttt{peter.elmer@cern.ch} \\
}

\begin{document}

\maketitle

\begin{abstract}
On-device machine learning is increasingly important in applications where extreme data rates or resource constraints make centralized processing infeasible. Logic neural networks (LNNs), which directly learn Boolean logic instead of conventional arithmetic operations, have recently emerged as a state-of-the-art approach for efficient inference and are particularly well suited for deployment in custom silicon. However, hard-wired solutions are typically optimized for narrowly defined tasks and lack the ability to adapt to changing data distributions.
Next-generation pixel detectors in particle physics present a particularly challenging use case: their high granularity and readout frequency generate data rates that cannot be centrally processed, while stringent space and power constraints necessitate highly efficient on-chip inference. Moreover, radiation damage progressively alters detector response, requiring models that can be re-trained and reconfigured after deployment.
In this work, we evaluate LNNs for on-chip inference in pixel detectors. We show that their superior resource efficiency may enable a step change in model capability, making more complex architectures feasible within the available hardware budget. LNNs also provide increased robustness against radiation-induced random bit flips. Finally, we present novel work on reconfigurable LNNs, enabling models to adapt to changing operating conditions after deployment.

\end{abstract}

\section{Introduction}

The increasing deployment of machine learning (ML) in real-world systems is driving inference closer to where data are generated. In many applications, transmitting raw data to a central processor is infeasible due to stringent requirements on latency, bandwidth, power, or privacy. This motivates on-device intelligence, where models perform inference directly on resource-constrained hardware. For example, always-on keyword spotting and wearable health monitoring operate on microwatt power budgets and kilobyte memory footprints~\cite{DBLP:journals/corr/abs-2106-07597}, autonomous driving systems require millisecond-scale decision making~\cite{lin2022low}, and data selection systems in particle physics experiments process data at rates approaching petabytes per second~\cite{aarrestad2021fast}.

Logic Neural Networks (LNNs) have recently demonstrated state-of-the-art tradeoffs between accuracy and computational footprint for small-scale inference~\cite{petersen2022deep,bacellar2024differentiable,petersen2024convolutional,andronic2025neuralut}. Rather than relying on arithmetic operations such as multiply-accumulate units and nonlinear activations, LNNs directly learn Boolean functions, such as AND and XOR, or higher-arity lookup tables (LUTs). This makes them particularly attractive for custom hardware, where learned logic can be directly mapped to digital circuits. However, the benefits of hard-wired logic also introduce a fundamental challenge: once deployed, the resulting circuit is specialized to the task and data distribution seen during training, potentially limiting its ability to adapt to changing conditions.

Future particle physics colliders provide a particularly demanding setting for adaptable on-device inference. At an event rate of 40~MHz, their pixel detectors will comprise billions of readout channels~\cite{CERN-LHCC-2017-009,CERN-LHCC-2017-021}, while detector response will evolve over time due to accumulated radiation damage. An on-device AI-based data preselection system must therefore not only be highly resource efficient, but also robust to hardware faults and capable of adapting after deployment. Recent advances in efficient ML architectures and hardware make such systems increasingly feasible, potentially enabling new approaches to data reduction and boosting the physics discovery potential of future detectors.

In this work, we use the publicly available SmartPixels datasets~\cite{morris_swartz_2024_10783560,shekar_2025_17180303} to investigate LNNs as an architecture for adaptive, resource-constrained inference in pixel detectors. We compare LNNs to quantized arithmetic-based neural networks, evaluating classification performance and hardware resource usage. We show that the superior inference efficiency of LNNs may enable a step change in model capability by making more complex architectures feasible within the available hardware budget. We further study their resilience to random bit flips, finding increased robustness. Finally, we present novel work on reconfigurable LNNs and investigate the associated tradeoff between reconfigurability and resource usage. Together, these results highlight LNNs as a promising building block for on-device intelligence in environments where efficiency, reliability, and adaptation must be considered jointly.

\section{Related Work}

The SmartPixels collaboration has conducted several studies using the public datasets considered in this work. A classifier for rejecting low-transverse-momentum ($p_T$) tracks has been developed and its deployment on an ASIC studied~\cite{Yoo_2024}. It separates high-$p_T$ tracks from positively and negatively charged low-$p_T$ tracks, exploiting the opposite bending of low-$p_T$ trajectories in the detector's magnetic field. Several quantized arithmetic architectures are considered, but only a simple dense model is synthesized for deployment using hls4ml~\cite{schulte2025hls4ml}; the convolutional architecture does not meet the hardware constraints. We use this dense quantized model as our baseline.

The effects of accumulated radiation damage on the SmartPixels system have also been studied~\cite{SHEKAR2026171897}. The resulting shift in the input distribution substantially degrades classifier performance, while retraining on the shifted data largely recovers the original accuracy. That work does not, however, consider the hardware implications of reconfigurability.

Neuromorphic computing has also been explored for the SmartPixels application, using spiking neural networks to replace quantized arithmetic with a fundamentally different computational paradigm~\cite{Kulkarni2023OnSensor}. Promising results have been demonstrated, but post-deployment adaptation is not addressed.

LNNs reach comparable efficiency while relying on mature digital hardware. Early approaches approximate lookup tables (LUTs) with parametrized continuous functions~\cite{susskind2022weightless,susskind2023uleen,andronic2023polylut,andronic2024neuralut,andronic2025neuralut,weng2025greater}, whereas more recent differentiable formulations learn LUTs directly and universally, covering both 2-input logic gates~\cite{petersen2022deep} and higher-arity LUTs~\cite{bacellar2024differentiable}. Such architectures achieve highly competitive accuracy--resource tradeoffs on small-scale classification tasks, but their suitability for adaptive, fault-tolerant on-device inference has received comparatively little attention.

Finally, quantization-aware training and sparsification are known to improve robustness to weight perturbations and bit-flip errors~\cite{He_2020_CVPR}. LNNs can be viewed as an extreme point in this design space, replacing arithmetic weights with discrete Boolean functions, and recent work shows empirically that this transition can yield a more favorable accuracy--resilience tradeoff for hardware fault tolerance~\cite{bacellar2026arithmetic}. We extend these observations to the SmartPixels application, studying robustness jointly with hardware efficiency and post-deployment reconfigurability.

\section{Experiments}

We evaluate dense QKeras~\cite{QKeras-2021}, dense LNN, and convolutional LNN classifiers on the SmartPixels benchmark. We use the same preprocessing of the raw charge images as~\cite{Yoo_2024}, giving 14 high-level input features for the dense models and a $13\times8$ time profile plus one high-level feature for the convolutional models; see \cref{app:data} for details.

All LNNs use exclusively 2-input LUTs (``logic gates'') with random, fixed connectivity, and a fixed binarization of the inputs by distributive thresholding into 100 ordered thresholds per feature. Following~\cite{petersen2022deep}, each gate is parametrized as a softmax-weighted sum over all 16 two-input Boolean functions and evaluated with fuzzy logic during training to make it differentiable; at inference the softmax is replaced by its argmax, yielding a discrete circuit. LNNs are implemented using \texttt{torchlogix}~\cite{torchlogix2026}. Exact architectures and training hyperparameters are given in \cref{app:models}.

As the intended deployment target is custom silicon, we use FPGA synthesis as a common surrogate for comparing hardware resource requirements. Since application-specific arithmetic can be implemented more efficiently as dedicated logic than through the general-purpose DSP blocks available on FPGAs, we disable DSP use and use the combined LUT+FF count as a logic-resource proxy. All models are synthesized for the Xilinx VU13P FPGA (\texttt{xcvu13p-flga2577-2-e}) with a requested clock period of 4.1667,ns (240,MHz). Trained LNNs are converted to RTL using \texttt{da4ml}~\cite{sun2025da4ml}, while QKeras models are converted using \texttt{hls4ml}~\cite{schulte2025hls4ml}. Input preprocessing and encoding, I/O and routing, and downstream decision logic are excluded from the reported post-synthesis resources; place-and-route and timing closure are not evaluated (left to future work)
\footnote{All source code (training, synthesis, and analysis) will be made publicly available if the paper is accepted.}.

\subsection{Hardware Efficiency and Classification Performance}
\label{sec:pareto}

\begin{figure}[h]
    \centering
    \includegraphics[width=\linewidth]{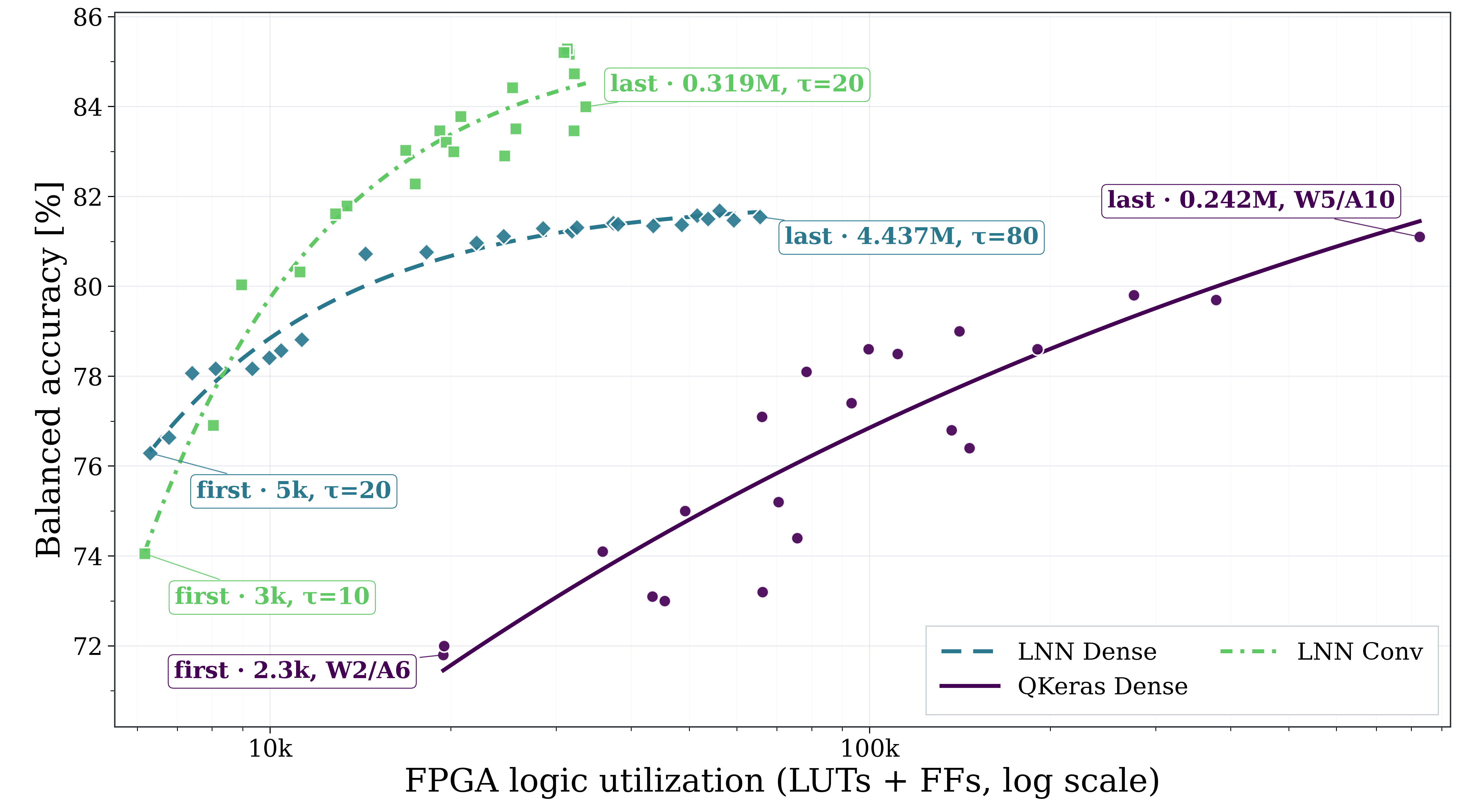}
    \caption{Balanced accuracy versus post-synthesis FPGA logic utilization for dense QKeras, dense LNN, and convolutional LNN architectures. Annotations identify the smallest and largest configurations in each sweep by number of trainable parameters, and report the \texttt{GroupSum} scaling parameter $\tau$ for LNNs and weight/activation precision (W/A) for QKeras.}
    \label{fig:resources}
\end{figure}

\Cref{fig:resources} compares balanced accuracy with post-synthesis FPGA logic utilization for dense QKeras, dense LNN, and convolutional LNN architectures. Dense LNNs provide a substantially better accuracy--resource tradeoff than the QKeras baselines. For example, at comparable balanced accuracy (0.807 for LNN and 0.798 for QKeras), the dense LNN requires 14,418 LUTs+FFs compared with 275,815 for QKeras, corresponding to approximately $19\times$ lower resource utilization. Across the architecture sweep, dense LNNs reach balanced accuracies above 0.81 while remaining in the tens-of-thousands resource regime.

The lower resource utilization of LNNs is particularly relevant for the time-resolved SmartPixels representation. Previous SmartPixels studies found that using eight charge profiles in a convolutional model improved performance over a dense single-profile model, but increased the parameter count by approximately $36\times$; the convolutional architecture was not pursued in the hardware implementation~\cite{Yoo_2024}. In contrast, our convolutional LNN operates on the same 105-input, eight-profile representation while requiring post-synthesis FPGA resources comparable to the dense LNNs and reaching balanced accuracies above 0.84.

\subsection{Robustness to Input Bit Flips}
\label{sec:bitflips}

Faults are applied to each model's digital input representation: 1,400 threshold-encoded bits for the LNN and 112 bits from the 8-bit representation of the 14 QKeras inputs. During fault-aware training, input bits are independently flipped with the specified probability on each forward pass. At inference, the same target \emph{fraction} of input bits is flipped at random; the reported fault rate therefore denotes the fraction of corrupted bits within each model's own input representation. Each nonzero evaluation point is averaged over five independent full-test realizations.

\begin{figure}[h]
    \centering
    \includegraphics[width=\linewidth]{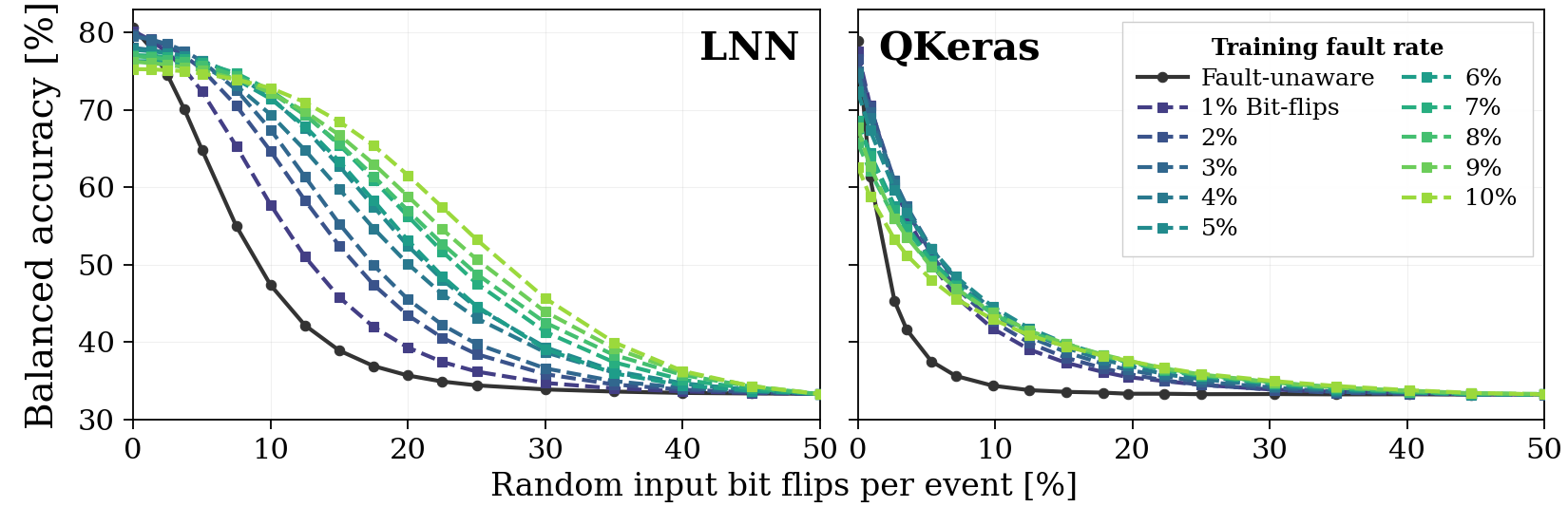}
    \caption{Balanced accuracy under random input-bit corruption for dense LNN (left) and QKeras (right) models with comparable fault-free performance.}
    \label{fig:bit-flips}
\end{figure}

We compare dense LNN and QKeras models with similar fault-free balanced accuracies of 0.807 and 0.798, respectively. \Cref{fig:bit-flips} shows how balanced accuracy changes as the fraction of corrupted input bits increases. Without fault-aware training, the LNN degrades substantially more slowly than QKeras. Training with random input bit flips improves robustness for both architectures, while the LNN maintains a pronounced advantage. At approximately 10\% input corruption, models trained without faults achieve approximately 47\% and 36\% balanced accuracy for LNN and QKeras, respectively; with 10\% fault-aware training they retain approximately 72\% and 46\%. Thus, at comparable fault-free accuracy, LNNs exhibit substantially greater robustness than QKeras to random input-bit corruption under the evaluated fault model.

\subsection{Reconfigurability}
\label{sec:reconf}
Radiation-induced sensor degradation progressively shifts detector response after deployment, requiring the on-chip classifier to be retrained and updated in the field~\cite{SHEKAR2026171897}. For conventional arithmetic models, this is straightforward: the computational graph is hard-wired once and the trained parameters are held in writable registers, so retraining only changes register contents. LNNs admit no such clean separation. Much of their efficiency comes from compactifying the Boolean computational graph at compile time, folding constants and merging redundant gates, so the surviving topology is itself a function of the trained truth tables. Retraining an LNN can therefore change its connectivity, not merely a set of stored coefficients.

To address this, we introduce a lightweight reconfigurability mechanism for LNNs. A designated subset of gates keeps its two-input connectivity fixed at synthesis time but reads its four-entry truth table from a small runtime configuration register instead of having it permanently wired; each such gate is realized as a multiplexer with two select bits and four configuration bits, stored in flip-flops, in place of a static logic gate. All remaining gates stay permanently fixed. This makes part of the deployed logic reprogrammable after fabrication without requiring general FPGA-style reconfigurable routing, at the cost of additional resources. That cost has a direct component: logic implementing the multiplexers and registers for the truth-tables, and an indirect component: the loss of compactification, since a reconfigurable gate can no longer be folded away.

Because the timing and severity of radiation-induced drift are only partially predictable, we evaluate reconfigurability under three assumed levels of prior knowledge about the data drift:
\begin{enumerate}[label=(\alph*)]
    \item \textbf{No knowledge:} the reconfigurable subset must be fixed before any exposure to the shifted data. Three approaches are considered: We either make entire layers reconfigurable, adding them from the output layer backwards (i), select the individual gates whose truth-table distribution is least converged after baseline training (ii), or compactify the baseline model first and retrain a reconfigurable model within that compactified topology (iii).
    \item \textbf{Exact knowledge:} an oracle in which the retrained model is known a priori and only the gates whose truth table changed are made reconfigurable. This gives the minimum resource overhead for a known target.
    \item \textbf{Partial knowledge:} assuming radiation effects can be simulated somewhat faithfully, we use one drifted operating condition to select a candidate reconfigurable subset, then retrain and evaluate that subset on the other drifted condition.
\end{enumerate}

\begin{figure}[h]
   \includegraphics[width=\linewidth]{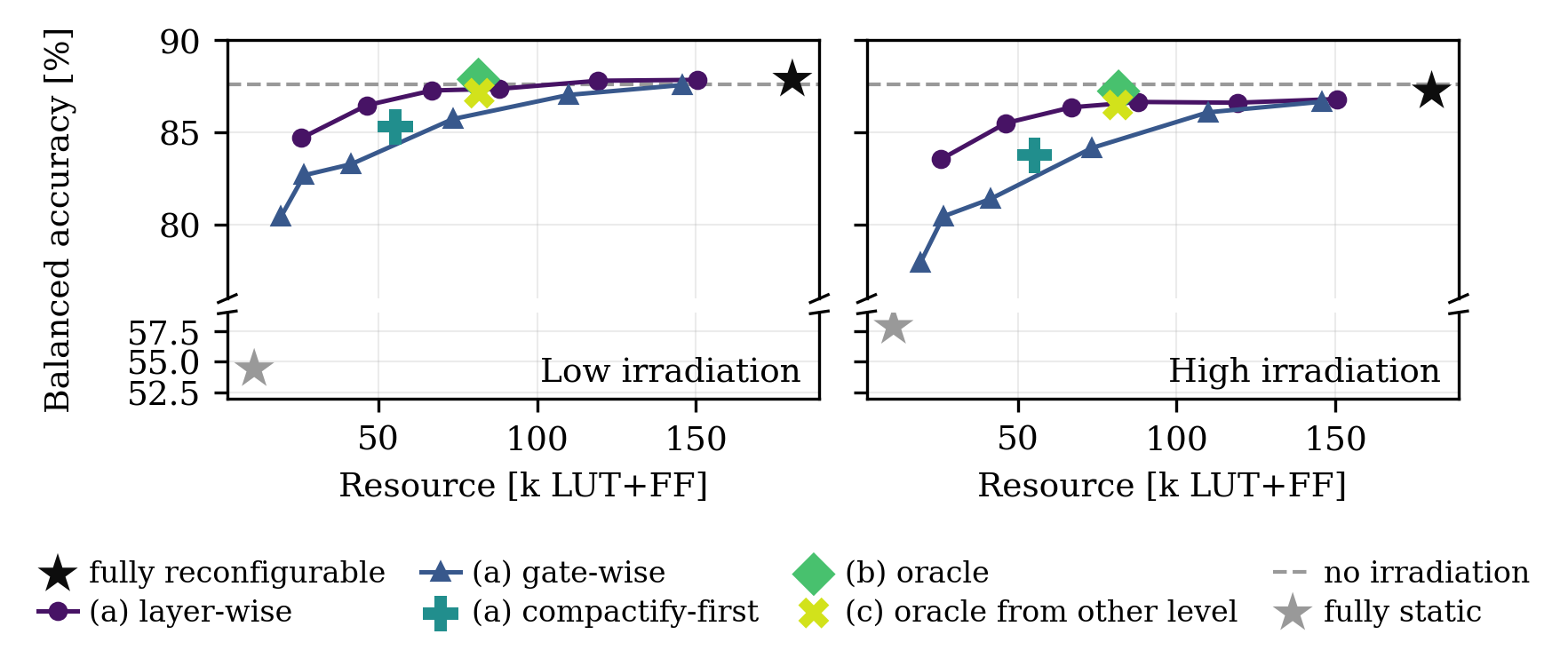}
    \caption{Balanced accuracy versus estimated resource usage for the baseline model retrained on the irradiated datasets at two accumulated fluences. No retraining (fully static) and unrestricted retraining (fully reconfigurable) are the two limiting cases. The remaining points use partial reconfigurability, with the reconfigurable subset chosen under different assumptions about a-priori knowledge of the data shift, indicated by the letters in parentheses.}
    \label{fig:reconf}
\end{figure}

We evaluate all three regimes on the dense LNN, trained on the non-irradiated SmartPixels sample and retrained on samples corresponding to two accumulated fluences (low and high)~\cite{shekar_2025_17180303}\footnote{This dataset differs slightly from the one used in \cref{sec:pareto,sec:bitflips}. Absolute numbers are not directly comparable.}. The resulting balanced accuracies are plotted against the reconfigurability-aware resource estimates in \cref{fig:reconf}. As expected, the fully static model degrades on the drifted data, while unrestricted retraining restores baseline performance. Making every gate reconfigurable, however, increases resource usage roughly 17-fold relative to the static model. The partially reconfigurable designs interpolate between these extremes. The oracle of scenario~\textbf{(b)} performs best among retrained models of comparable size, as expected. Scenario~\textbf{(c)}, which selects the reconfigurable subset on one irradiation scenario and applies it to the other, offers no advantage over scenario~\textbf{(a)}, which assumes no knowledge of the drift at all. The latter performs surprisingly well: with three reconfigurable layers it retains 99\% of the fully retrained accuracy at 37\% of the resource cost of full reconfigurability. Notably, making whole layers reconfigurable Pareto-dominates the gate-wise criteria\footnote{We evaluated several strategies for choosing which gates to make reconfigurable and report only the best-performing one.}.

\section{Limitations and Future Work}

The hardware results in this work use FPGA synthesis as a proxy for on-chip implementation cost; future work will require ASIC synthesis, place-and-route, and power analysis for more realistic deployment estimates. A first SoC integrating an LNN on an eFPGA is targeted for tape-out in 2027. The robustness studies currently consider input bit flips only, and should be extended to faults in intermediate activations and to convolutional LNNs. Reconfigurability is likewise evaluated only for dense models. Lastly, further optimizations can be applied to LNNs, e.g., by learning the connectivity~\cite{mommen2025method} and/or the binarization thresholds~\cite{gerlach2025warp}. In particular the latter could improve bit-efficiency when including the cost of data I/O in future works.

\begin{ack}
We thank the SmartPixels Collaboration for their valuable discussions, feedback, and continued input throughout this work. We also thank members of the MAIA detector and muon-collider beam-induced-background mitigation efforts for helpful discussions on detector operating conditions and radiation effects.

This work was supported by the National Science Foundation under Cooperative Agreements OAC-1836650 and PHY-2323298.
\end{ack}

\printbibliography

@article{bacellar2026arithmetic,
  author  = {Bacellar, Alan T. L. and
             Chemudupati, Sathvik and
             Nag, Shashank and
             Seigler, Allison and
             Lima, Priscila Machado Vieira and
             Fran{\c{c}}a, Felipe M. G. and
             John, Lizy Kurian},
  title   = {From Arithmetic to Logic: The Resilience of Logic and Lookup-Based Neural Networks Under Parameter Bit-Flips},
  journal = {Transactions on Machine Learning Research},
  year    = {2026},
  month   = jul,
}

@article{Yoo_2024,
    doi = {10.1088/2632-2153/ad6a00},
    url = {https://doi.org/10.1088/2632-2153/ad6a00},
    year = {2024},
    month = {aug},
    publisher = {IOP Publishing},
    volume = {5},
    number = {3},
    pages = {035047},
    author = {Yoo, Jieun and Dickinson, Jennet and Swartz, Morris and Di Guglielmo, Giuseppe and Bean, Alice and Berry, Douglas and Blanco Valentin, Manuel and DiPetrillo, Karri and Fahim, Farah and Gray, Lindsey and Hirschauer, James and Kulkarni, Shruti R and Lipton, Ron and Maksimovic, Petar and Mills, Corrinne and Neubauer, Mark S and Parpillon, Benjamin and Pradhan, Gauri and Syal, Chinar and Tran, Nhan and Wen, Dahai and Young, Aaron},
    title = {Smart pixel sensors: towards on-sensor filtering of pixel clusters with deep learning},
    journal = {Machine Learning: Science and Technology},
}

@article{SHEKAR2026171897,
    title = {Robustness of the smartpixels classifier for different simulated sensor geometries and non-ideal detector conditions},
    journal = {Nuclear Instruments and Methods in Physics Research Section A: Accelerators, Spectrometers, Detectors and Associated Equipment},
    volume = {1092},
    pages = {171897},
    year = {2026},
    issn = {0168-9002},
    doi = {https://doi.org/10.1016/j.nima.2026.171897},
    url = {https://www.sciencedirect.com/science/article/pii/S0168900226006236},
    author = {Danush Shekar and Ben Weiss and Morris Swartz and Corrinne Mills and Jennet Dickinson and Lindsey Gray and David Jiang and Mohammad Abrar Wadud and Daniel Abadjiev and Anthony Badea and Douglas Berry and Alec Cauper and Arghya Ranjan Das and Karri Folan DiPetrillo and Farah Fahim and Rachel {Kovach Fuentes} and Abhijith Gandrakota and Giuseppe {Di Guglielmo} and Eliza Howard and Shiqi Kuang and Carissa Kumar and Mia Liu and Petar Maksimovic and Nick Manganelli and Mark S. Neubauer and Aidan Nicholas and Emily Pan and Benjamin Parpillon and Jannicke Pearkes and Ricardo Silvestre and Chinar Syal and Amit Trivedi and Keith Ulmer and Jieun Yoo and Eric You},
    keywords = {Edge computing, On-chip networks, Smart detectors, Data-reduction, Detector simulation},
    % abstract = {Pixel tracking detectors at upcoming collider experiments will see unprecedented charged-particle densities. Real-time data reduction on the detector will enable higher granularity and faster readout, possibly enabling the use of the pixel detector in high-rate online event selection, such as the ATLAS or CMS first-level trigger systems. This data reduction can be accomplished with a neural network (NN) in the readout chip bonded with the sensor that recognizes and rejects tracks with low transverse momentum (pT) based on the geometrical shape of the charge deposition (“cluster”). To design viable detectors for deployment, the dependence of the NN as a function of the sensor geometry, external magnetic field, irradiation, and noise must be understood. In this paper, we present first studies of the efficiency and data reduction for planar pixel sensors exploring these parameters. For the CMS HL-LHC sensor geometry, we obtain a signal efficiency of (91.9 ± 0.7)% and a data reduction of (29.7 ± 1.0)%. A smaller sensor pitch in the bending direction improves the pT discrimination, but a larger pitch can be partially compensated with detector thickness. Any accumulated radiation damage also changes the cluster shape, reducing the signal efficiency compared to the baseline by approximately 30–60% in absolute terms, but nearly all of the performance can be recovered through retraining of the network and updating the weights. Finally, the impact of noise was investigated, and retraining the network on noise-injected datasets was found to maintain performance within 6% of the baseline network trained and evaluated on noiseless data.}
}

@inproceedings{Kulkarni2023OnSensor,
  author    = {Kulkarni, Shruti R. and
               Young, Aaron and
               Date, Prasanna and
               Miniskar, Narasinga Rao and
               Vetter, Jeffrey S. and
               Fahim, Farah and
               Parpillon, Benjamin and
               Dickinson, Jennet and
               Tran, Nhan and
               Yoo, Jieun and
               Mills, Corrinne and
               Swartz, Morris and
               Maksimovic, Petar and
               Schuman, Catherine and
               Bean, Alice},
  title     = {On-Sensor Data Filtering using Neuromorphic Computing for High Energy Physics Experiments},
  booktitle = {Proceedings of the 2023 International Conference on Neuromorphic Systems},
  series    = {ICONS '23},
  year      = {2023},
  publisher = {Association for Computing Machinery},
  address   = {Santa Fe, NM, USA},
  isbn      = {9798400701757},
  doi       = {10.1145/3589737.3605976},
  url       = {https://doi.org/10.1145/3589737.3605976},
  pages     = {1--8}
}

@inproceedings{susskind2022weightless,
  title={Weightless neural networks for efficient edge inference},
  author={Susskind, Zachary and Arora, Aman and Miranda, Igor DS and Villon, Luis AQ and Katopodis, Rafael F and De Ara{\'u}jo, Leandro S and Dutra, Diego LC and Lima, Priscila MV and Fran{\c{c}}a, Felipe MG and Breternitz Jr, Mauricio and others},
  booktitle={Proceedings of the 31st International Conference on Parallel Architectures and Compilation Techniques (PACT)},
  pages={279--290},
  year={2022}
}

@article{susskind2023uleen,
  title={Uleen: A novel architecture for ultra-low-energy edge neural networks},
  author={Susskind, Zachary and Arora, Aman and Miranda, Igor DS and Bacellar, Alan TL and Villon, Luis AQ and Katopodis, Rafael F and De Ara{\'u}jo, Leandro S and Dutra, Diego LC and Lima, Priscila MV and Fran{\c{c}}a, Felipe MG and others},
  journal={Transactions on Architecture and Code Optimization (TACO)},
  volume={20},
  number={4},
  pages={1--24},
  year={2023},
  publisher={ACM}
}

@inproceedings{andronic2023polylut,
  title={PolyLUT: learning piecewise polynomials for ultra-low latency FPGA LUT-based inference},
  author={Andronic, Marta and Constantinides, George A},
  booktitle={Proceedings of the 2023 International Conference on Field Programmable Technology (ICFPT)},
  pages={60--68},
  year={2023},
  organization={IEEE}
}

@inproceedings{andronic2024neuralut,
  title={NeuraLUT: Hiding neural network density in boolean synthesizable functions},
  author={Andronic, Marta and Constantinides, George A},
  booktitle={Proceedings of the 34th International Conference on Field-Programmable Logic and Applications (FPL)},
  pages={140--148},
  year={2024},
  organization={IEEE}
}

@inproceedings{andronic2025neuralut,
  title={NeuraLUT-Assemble: Hardware-aware Assembling of Sub-Neural Networks for Efficient LUT Inference},
  author={Andronic, Marta and Constantinides, George A},
  booktitle={Proceedings of the 33rd IEEE Annual International Symposium on Field-Programmable Custom Computing Machines (FCCM)},
  pages={208--216},
  year={2025},
  organization={IEEE}
}

@inproceedings{weng2025greater,
  title={Greater than the Sum of its LUTs: Scaling Up LUT-based Neural Networks with AmigoLUT},
  author={Weng, Olivia and Andronic, Marta and Zuberi, Danial and Chen, Jiaqing and Geniesse, Caleb and Constantinides, George A and Tran, Nhan and Fraser, Nicholas J and Duarte, Javier Mauricio and Kastner, Ryan},
  booktitle={Proceedings of the 2025 ACM/SIGDA International Symposium on Field Programmable Gate Arrays},
  pages={25--35},
  year={2025}
}

@article{petersen2022deep,
  title={Deep differentiable logic gate networks},
  author={Petersen, Felix and Borgelt, Christian and Kuehne, Hilde and Deussen, Oliver},
  journal={Advances in Neural Information Processing Systems},
  volume={35},
  pages={2006--2018},
  year={2022}
}

@inproceedings{mommen2025method,
  title={A Method for Optimizing Connections in Differentiable Logic Gate Networks},
  author={Mommen, Wout},
  booktitle={International Conference on Neuromorphic Computing Systems 2025},
  year={2025}
}

@inproceedings{bacellar2024differentiable,
  title={Differentiable weightless neural networks},
  author={Bacellar, Alan TL and Susskind, Zachary and Breternitz, Maur{\'\i}cio and John, Eugene and John, Lizy K and Lima, Priscila MV and Fran{\c{c}}a, Felipe MG and others},
  booktitle={Proceedings of the 41st International Conference on Machine Learning, PMLR},
  volume={235},
  pages={2277--2295},
  year={2024}
}

@article{petersen2024convolutional,
  title={Convolutional differentiable logic gate networks},
  author={Petersen, Felix and Kuehne, Hilde and Borgelt, Christian and Welzel, Julian and Ermon, Stefano},
  journal={Advances in Neural Information Processing Systems},
  volume={37},
  pages={121185--121203},
  year={2024}
}

@article{sun2025da4ml,
  title={da4ml: Distributed Arithmetic for Real-time Neural Networks on FPGAs},
  author={Sun, Chang and Que, Zhiqiang and Loncar, Vladimir and Luk, Wayne and Spiropulu, Maria},
  journal={arXiv preprint arXiv:2507.04535},
  year={2025}
}

@article{lin2022low,
  title={Low-latency edge video analytics for on-road perception of autonomous ground vehicles},
  author={Lin, Jie and Yang, Peng and Zhang, Ning and Lyu, Feng and Chen, Xianfu and Yu, Li},
  journal={IEEE Transactions on Industrial Informatics},
  volume={19},
  number={2},
  pages={1512--1523},
  year={2022},
  publisher={IEEE}
}

@article{aarrestad2021fast,
  title={Fast convolutional neural networks on FPGAs with hls4ml},
  author={Aarrestad, Thea and Loncar, Vladimir and Ghielmetti, Nicol{\`o} and Pierini, Maurizio and Summers, Sioni and Ngadiuba, Jennifer and Petersson, Christoffer and Linander, Hampus and Iiyama, Yutaro and Di Guglielmo, Giuseppe and others},
  journal={Machine Learning: Science and Technology},
  volume={2},
  number={4},
  pages={045015},
  year={2021},
  publisher={IOP Publishing}
}

@article{gerlach2025warp,
  title={WARP-LUTs--Walsh-Assisted Relaxation for Probabilistic Look Up Tables},
  author={Gerlach, Lino and V{\aa}ge, Liv and Gerlach, Thore and Kauffman, Elliott and Ojalvo, Isobel},
  journal={arXiv preprint arXiv:2510.15655},
  year={2025}
}

@article{schulte2025hls4ml,
  title={hls4ml: A Flexible, Open-Source Platform for Deep Learning Acceleration on Reconfigurable Hardware},
  author={Schulte, Jan-Frederik and Ramhorst, Benjamin and Sun, Chang and Mitrevski, Jovan and Ghielmetti, Nicol{\`o} and Lupi, Enrico and Danopoulos, Dimitrios and Loncar, Vladimir and Duarte, Javier and Burnette, David and others},
  journal={arXiv preprint arXiv:2512.01463},
  year={2025}
}

@article{DBLP:journals/corr/abs-2106-07597,
  author       = {Colby R. Banbury and
                  Vijay Janapa Reddi and
                  Peter Torelli and
                  Jeremy Holleman and
                  Nat Jeffries and
                  Csaba Kir{\'{a}}ly and
                  Pietro Montino and
                  David Kanter and
                  Sebastian Ahmed and
                  Danilo Pau and
                  Urmish Thakker and
                  Antonio Torrini and
                  Pete Warden and
                  Jay Cordaro and
                  Giuseppe Di Guglielmo and
                  Javier M. Duarte and
                  Stephen Gibellini and
                  Videet Parekh and
                  Honson Tran and
                  Nhan Tran and
                  Wenxu Niu and
                  Xuesong Xu},
  title        = {MLPerf Tiny Benchmark},
  journal      = {CoRR},
  volume       = {abs/2106.07597},
  year         = {2021},
  url          = {https://arxiv.org/abs/2106.07597},
  eprinttype   = {arXiv},
  eprint       = {2106.07597},
  bibsource    = {dblp computer science bibliography, https://dblp.org}
}

@software{torchlogix2026,
  author = {Gerlach, Lino and
            Gerlach, Thore and
            Kauffman, Elliott and
            V{\aa}ge, Liv},
  title = {torchlogix},
  year = {2026},
  doi = {10.5281/zenodo.18800427},
  url = {https://github.com/ligerlac/torchlogix}
}

@article{QKeras-2021, author = {Coelho, Claudionor N. Jr. and Kuusela, Aki and Li, Shan and Zhuang, Hao and Ngadiuba, Jennifer and Aarrestad, Thea Klaeboe and Loncar, Vladimir and Pierini, Maurizio and Pol, Adrian Alan and Summers, Sioni}, title = {Automatic heterogeneous quantization of deep neural networks for low-latency inference on the edge for particle detectors}, journal = {Nature Machine Intelligence}, volume = {3}, pages = {675--686}, year = {2021}, doi = {10.1038/s42256-021-00356-5} }

@dataset{morris_swartz_2024_10783560,
  author       = {Morris Swartz and
                  Jennet Dickinson},
  title        = {Smart pixel dataset},
  month        = mar,
  year         = 2024,
  publisher    = {Zenodo},
  version      = 3,
  doi          = {10.5281/zenodo.10783560},
  url          = {https://doi.org/10.5281/zenodo.10783560},
}

@dataset{shekar_2025_17180303,
  author       = {Shekar, Danush and
                  Swartz, Morris and
                  Dickinson, Jennet and
                  Wadud, Mohammad Abrar},
  title        = {Smartpixels co-design datasets},
  month        = sep,
  year         = 2025,
  publisher    = {Zenodo},
  version      = 1,
  doi          = {10.5281/zenodo.17180303},
  url          = {https://doi.org/10.5281/zenodo.17180303},
}

@techreport{CERN-LHCC-2017-021,
      collaboration = "ATLAS",
      author        = "The ATLAS Collaboration",
      title         = "{Technical Design Report for the ATLAS Inner Tracker Pixel
                       Detector}",
      institution   = "CERN",
      reportNumber  = "CERN-LHCC-2017-021, ATLAS-TDR-030",
      address       = "Geneva",
      year          = "2017",
      url           = "https://cds.cern.ch/record/2285585",
      doi           = "10.17181/CERN.FOZZ.ZP3Q",
}

@techreport{CERN-LHCC-2017-009,
      collaboration = "CMS",
      author        = "The CMS Collaboration",
      title         = "{The Phase-2 Upgrade of the CMS Tracker}",
      institution   = "CERN",
      reportNumber  = "CERN-LHCC-2017-009, CMS-TDR-014",
      address       = "Geneva",
      year          = "2017",
      url           = "https://cds.cern.ch/record/2272264",
      doi           = "10.17181/CERN.QZ28.FLHW",
}

@InProceedings{He_2020_CVPR,
author = {He, Zhezhi and Rakin, Adnan Siraj and Li, Jingtao and Chakrabarti, Chaitali and Fan, Deliang},
title = {Defending and Harnessing the Bit-Flip Based Adversarial Weight Attack},
booktitle = {Proceedings of the IEEE/CVF Conference on Computer Vision and Pattern Recognition (CVPR)},
month = {June},
year = {2020}
}

\appendix
\crefalias{section}{appendix}
\crefalias{subsection}{appendix}

\section{Data Structure and Preprocessing}
\label{app:data}

Each SmartPixels event consists of 20 time samples of a $13\times21$ pixel charge image, together with the local track position $y_0$. We follow the preprocessing of~\cite{Yoo_2024}.

For the \textbf{dense} models, only the final time sample is used. The charge image is projected along the $x$ dimension by averaging, giving a 13-component charge profile; together with $y_0$ this yields the 14 input features.

For the \textbf{convolutional} models, the same projection is applied to the first eight time samples, giving a $13\times8$ time-profile representation. Together with $y_0$ this yields 105 inputs in total.

Charge values are transformed with a symmetric-log scaling before being passed to the models. The task is a three-class problem, separating positively charged low-$p_T$, negatively charged low-$p_T$, and high-$p_T$ tracks. Training uses an inverse-class-frequency weighted cross-entropy loss and we report balanced accuracy throughout. Train and test samples are drawn from disjoint sets of data files, following an 80/20 train/test split: 128 training and 32 test files for the dataset used in \cref{sec:pareto,sec:bitflips}, and 16 and 4 for the irradiated dataset used in \cref{sec:reconf}. A further 1\% of the training files is held out for validation. All splits use a fixed seed of 42.

\section{Model Architectures and Training}
\label{app:models}

\paragraph{Dense LNNs.}
The dense LNN sweep of \cref{sec:pareto} combines a three-layer family for the smallest models and a seven-layer family for the larger models, with layer widths varied to sweep model size. Writing the layer widths as a tuple, the smallest configuration is $(1269,1269,1269)$ and the largest is $(33600,33600,33600,22400,22400,22400,16800)$. The 14 input features are binarized into $14\times100=1400$ bits, and the final layer feeds a \texttt{GroupSum} over three groups with scaling parameter $\tau$. Across the evaluated sweep, $\tau$ takes values in $\{5,10,20,30,40,60,80\}$. Every \texttt{LogicDense} layer uses \texttt{lut\_rank}\, $=2$ (i.e.\ two-input LUTs) and residual weight initialization.

The reconfigurability study of \cref{sec:reconf} uses a single dense model of shape $(6300,6300,6300,4200,4200,4200,3150)$ with $\tau=20$: seven logic layers, resulting in 554,000 trainable parameters.

\paragraph{Convolutional LNN.}
The convolutional LNN sweep uses a common sequence of four \texttt{LogicConv2d} layers, one \texttt{OrPooling2d} stage, and four \texttt{LogicDense} layers, while the kernel counts and dense-layer widths are varied to sweep model size. The convolutional layers use $3\times3$ receptive fields, tree depth 3, and \texttt{lut\_rank}\,$=2$. In the reference configuration, the first two convolutional layers contain 256 and 512 kernels, followed by an \texttt{OrPooling2d} stage with kernel size and stride 2 and two further convolutional layers with 1024 and 960 kernels. The flattened output is concatenated with the 100 binarized $y_0$ bits and passed through four \texttt{LogicDense} layers of widths $(12000,12000,6000,3000)$ into a \texttt{GroupSum} over three groups.

\paragraph{QKeras baselines.}
The quantized baselines follow the dense Model-2 architecture of~\cite{Yoo_2024}: fully connected layers on the 14 input features, with batch normalization after the first layer and quantized ReLU activations. The plotted sweep contains 21 trained and synthesized configurations. Five weight/ activation precision settings---W2/A4, W2/A6, W3/A6, W4/A8, and W5/A10---are represented for two variants of the $(128)$ topology and for the $(192)$ and $(256,64)$ topologies. The $(768,256,128)$ topology is included at W5/A10.

\paragraph{Trend fits.}
For each model family, balanced accuracy is fitted independently as a saturating function of FPGA resource utilization,
  \[
  A(R)=A_{\infty}
  +c\exp\!\left[
  -\frac{\log_{10}R-\log_{10}R_{\min}}{s}
  \right],
  \qquad c\leq0,\quad s>0,
  \]
where $R=\mathrm{LUT}+\mathrm{FF}$. The saturation scale $s$ is scanned over 450 logarithmically spaced values. At each value, $A_{\infty}$ and $c$ are determined by unweighted least squares, and the solution with the smallest mean squared residual is retained. Each fitted curve is evaluated only over the resource range spanned by its corresponding models.

\paragraph{Training.}
All LNNs are trained with AdamW (learning rate $0.01$, no weight decay), batch size 1024, and a class-weighted cross-entropy loss, with a fixed seed of 42. The sweep of \cref{sec:pareto} trains each configuration for at least 20{,}000 steps and at least five hours of wall-clock time; the reconfigurability study trains for at least 30{,}000 steps and at least two hours, after which balanced accuracy was observed to have plateaued. Models are evaluated every 5{,}000 steps and the best checkpoint by validation balanced accuracy is retained.

\end{document}